\documentclass[final,3p,times,12pt]{elsarticle}

\usepackage{graphicx}
\usepackage[bookmarks=false]{hyperref}
\usepackage[pdftex]{color}
\usepackage[font=footnotesize,labelfont=bf]{caption}
\usepackage[font=footnotesize,labelfont=bf]{subfig}

\usepackage{standalone}
\usepackage{booktabs}
\usepackage{anyfontsize}
\usepackage{svg}

\usepackage[utf8]{inputenc}
\usepackage{amsmath}
\usepackage{amsopn}
\usepackage{accents}
\usepackage{siunitx}
\usepackage{amssymb}
\usepackage{pifont}
\usepackage{mathrsfs} 
\usepackage{eucal} 
\renewcommand{\mathcal}[1]{\CMcal{#1}} 
\usepackage{trfsigns} 
\usepackage{nicefrac} 
\usepackage{url}

\usepackage{mleftright} 
\mleftright 
\usepackage{array} 

\newcommand{\Reals}{\mathbb{R}}      

\DeclareUnicodeCharacter{2212}{-}

\usepackage{lineno}

\makeatletter
\def\ps@pprintTitle{%
 \let\@oddhead\@empty
 \let\@evenhead\@empty
 \def\@oddfoot{\centerline{\thepage}}%
 \let\@evenfoot\@oddfoot}
\makeatother
\journal{Neurocomputing}

\begin{document}

\begin{frontmatter}


\title{Exploring Embedding Methods in Binary Hyperdimensional Computing: A Case Study for Motor-Imagery based Brain--Computer Interfaces}



\author[eth]{Michael Hersche}
\author[epfl]{José del R. Millán}
\author[eth]{Luca Benini}
\author[eth]{Abbas Rahimi}
\address[eth]{Department of Information Technology and Electrical Engineering, ETH Zurich, Zürich, Switzerland}
\address[epfl]{Defitech Foundation Chair in Brain-Machine Interface, EPFL, Lausanne, Switzerland}

\begin{abstract}
Key properties of brain-inspired hyperdimensional (HD) computing make it a prime candidate for energy-efficient and fast learning in biosignal processing.
The main challenge is however to formulate embedding methods that map biosignal measures to a binary HD space.
In this paper, we explore variety of such embedding methods and examine them with a challenging application of motor imagery brain--computer interface (MI-BCI) from electroencephalography (EEG) recordings.
We explore embedding methods including random projections, quantization based thermometer and Gray coding, and learning HD representations using end-to-end training.
All these methods, differing in complexity, aim to represent EEG signals in binary HD space, e.g. with 10,000 bits.  
%
%
%
This leads to development of a set of HD learning and classification methods that can be selectively chosen (or configured) based on accuracy and/or computational complexity requirements of a given task.
We compare them with state-of-the-art linear support vector machine (SVM) on an NVIDIA TX2 board using the 4-class BCI competition IV-2a dataset as well as a new 3-class dataset.
Compared to SVM, results on 3-class dataset show that simple thermometer embedding achieves moderate average accuracy (79.56\% vs. 82.67\%) with 26.8$\times$ faster training time and 22.3$\times$ lower energy; on the other hand, switching to end-to-end training with learned HD representations wipes out these training benefits while boosting the accuracy to 84.22\% (1.55\% higher than SVM).
Similar trend is observed on the 4-class dataset where SVM achieves on average 74.29\%: the thermometer embedding achieves 89.9$\times$ faster training time and 58.7$\times$ lower energy, but a lower accuracy (67.09\%) than the learned representation of 72.54\%. 
%
%
\end{abstract}


\end{frontmatter}


\section{Introduction}\label{sec:intro}
%
Hyperdimensional (HD) computing~\cite{KanervaPentti2009HCAI} is based on the intuition that brains compute with patterns of neural activity that are not readily associated with scalar numbers.
Due to the very large size of the brain’s circuits (fan-ins and fan-outs can be in tens of thousands), we can model neural activity patterns with points of an HD space,
that is, with \emph{hypervectors.}
When the dimensionality is in the thousands~\cite{Kanerva1988}, operations with hypervectors create a computational behavior with unique features in terms of robustness, energy-efficiency, speed of learning, etc.~\cite{Rahimi2017_nanoscale}.
These properties make HD computing a prime candidate for 
applications related to how humans interact with the world around them and with the cyberworld through ``wearable'' devices for which biosignal training datasets are small, individual variability is significant, privacy, latency, and energy efficiency demands are tight~\cite{Proc18_ExG}.

Such application domain includes brain--computer interfaces (BCIs) that enable a communication channel between a user and an external device through intentional modulation of brain signals, e.g., motor imagery (MI) movement of a part of body~\cite{BCI_Neurocomputing17}.  
A BCI aims to recognize human intentions from the analysis of spatiotemporal neural activity, typically recorded non-invasively by a number of electroencephalogram (EEG) electrodes.
Such information can enable control games~\cite{Saeedi2016,Perdikis2018}, drive a wheelchair~\cite{Carlson_Millan2013}, and even motor rehabilitation after stroke~\cite{Ramos-Murguialday2013, biasiucci2018brain}.
Despite successful enhancements using machine learning techniques, BCIs are still facing challenges regarding the classification accuracy, interpretability, explainability, and usability for online deployment of the system~\cite{SchirrmeisterRobinTibor2017Dlwc}.
The difficulty in determining human intentions reliably is mostly due to the low signal-to-noise ratio (SNR) of the EEG signal and the high variability among subjects and over time~\cite{Grosse-Wentrup2013,Saeedi2016}.  
These may be addressed by a computational paradigm such as HD computing that exhibits robustness against low SNR conditions, supports fast and incremental learning (hence they are easy to calibrate), and have very energy-efficient implementation.
However, one main challenge is to realize efficient \emph{embedding} methods that map EEG measurements directly to hypervectors.

Principles of HD computing are applied to create systems capable of solving cognitive tasks, for example, 
Raven's Progressive Matrices~\cite{BSC+SDM}, 
analogical reasoning~\cite{Analog_Map_2012,Dollar-Mexico,Analogy_Plate2000},
robust memory retrieval~\cite{ALSTM,HD_Capacity_NC18},
robot learning by demonstration~\cite{Robot_LearnBehHierarchies,Robot_LearnByDem},
and several learning and classification tasks robust at low SNR conditions~\cite{JSSC18,Rahimi2016,SparseHD_TNNLS18,Rahimi2017_nanoscale}.
We speak of high-dimensionality when the dimension of the hypervector is in the order of thousands~\cite{Kanerva1988}, e.g. 10,000-$d$. 
These hypervectors are \emph{holographic} and (pseudo)random with independent and identically distributed (i.i.d.) components. 
The components of hypervectors can be
binary~\cite{Kanerva1996},
bipolar~\cite{VSA03}, 
integer~\cite{Integer_Echo_State}, 
real~\cite{Real_Com_Bin_HD12,KellyMatthewA2013}, or 
complex~\cite{PlateT.A.1995Hrr,PlateBook}.
These different frameworks are in fact related to each other.
Dense binary (equally probable $0$s and $1$s) is mathematically
equivalent to the bipolar.
The real and the complex are related via Fourier transform, and the bipolar is equivalent to the complex when the phase angles are restricted to 0 and $\pi$~\cite{PlateBook}.
To keep representations and operations simple, we focus only on dense binary hypervectors, aka binary spatter coding~\cite{Kanerva1996}.

For learning and classification tasks, HD computing has been initially applied to text analytics where each discrete symbol can be readily mapped to a hypervector~\cite{Randomindexing,Aditya,Rahimi2016,Rasti2016,Reasoning_with_vectors}.
More recently, it has been extended to operate with multichannel analog inputs in several biosignal processing applications~\cite{BioCAS18,RahimiAbbas2016HbpA,ISCAS18,DAc18,RahimiAbbas2017HCfB,BICT17,Proc18_ExG}.
HD computing linearly scales with the number of electrodes and maintains its accuracy with various types of biosignal acquisitions: ranging from intracranial EEG (or ECoG) electrodes with the highest SNR~\cite{BioCAS18}, to electromyography (EMG) electrodes with relatively lower SNR~\cite{RahimiAbbas2016HbpA,ISCAS18,DAc18}, and finally to the EEG electrodes with the lowest SNR for event related potential (ERP) detection~\cite{RahimiAbbas2017HCfB,BICT17}; (see~\cite{Proc18_ExG} for an overview). 
%
%
However, there is a lack of universal methods that allow to map arbitrary number of real-valued features to such binary HD space.

The main contribution of this paper is to develop a set of universally applicable encoding schemes to represent EEG signals in binary HD space. 
We first propose to map real-valued features effectively to $d$-dimensional binary hypervectors with embedding methods such as random projections, and quantization-based thermometer and Gray coding.
Random projections are then extended to \textit{learned projections} by training the projections in order to produce most distinctive hypervectors.  
Inspired by binary neural networks (BNNs)~\cite{MontoneGuglielmo2017Hcfa}, our encoder is trained end-to-end using gradient descent with backpropagation.
Learned projections prove to be much more effective in terms of classification accuracy as well as required dimensionality. 
The aforementioned mappings are evaluated on multi-spectral real-valued Riemannian features from the 4-class MI-dataset of the BCI competition IV-2a~\cite{BCI-recording} as well as a new three class MI-dataset based on experiments in~\cite{Saeedi2016}.
We also provide open access to that dataset and source code\footnote{Source code and 3-class dataset are available at \url{https://github.com/MHersche/HDembedding-BCI}}.
Several configurations are compared with state-of-the-art linear SVM in terms of classification accuracy, memory footprint as well as power consumption and run time in training and inference on a NVIDIA TX2 board yielding to the following main results: 
\begin{itemize}
    \item On the 3-class dataset, thermometer embedding achieve moderate average classification accuracy compared to the SVM (79.56\% vs. 82.67\%) with 26.8$\times$ faster training time and 22.3$\times$ lower energy consumption in training.  
    \item Similar trend is observed on the 4-class dataset where SVM achieves on  average 74.29\% and the thermometer embedding 67.09\% with 89.0$\times$ faster training time and 58.7$\times$ lower energy energy consumption in training.
    \item On both datasets Gray embedding as well as random projection perform similar to thermometer embedding, however, when using learned projections the average accuracy increases to 84.22\% on the 3-class dataset, which is 1.55\% higher than SVM, and 72.54\% on the 4-class dataset.
    \item Learned projections demonstrate to be much more efficient than random projections requiring a lower dimensionality by order of magnitude at higher average classification accuracy. However, the boost in classification accuracy comes with additional cost in training compared to random and quantization-based embeddings.
\end{itemize}

The rest of this paper is organized as follows. 
In Section \ref{sec:hdcomp}, we introduce the Riemannian feature extraction method as well as HD computing and show their application in MI-BCIs. 
In Section \ref{sec:hd_mapping}, we present our main contribution by proposing a new encoder which maps real-valued features to binary hypervectors. 
Our experimental results are described in Section \ref{sec:results} followed by discussion in Section \ref{sec:disc}. Section \ref{sec:conclusion} concludes the paper.   

\section{Background}\label{sec:hdcomp}
\subsection{Riemannian Covariance  Features}\label{subsec:feat_extr}
Traditional BCIs extract features from the EEG signal using spectral-power features in connection with a spatial filter, which is learned to maximize the discriminability between two classes.
This is better known as filter bank common spatial pattern (FBCSP)~\cite{Kai_2008_winner_bci}.
Here we use a more recent approach which is to directly manipulate spatial EEG covariance matrices using the dedicated Riemannian geometry~\cite{Lotte_Riemann2017}.
In contrast with CSP, this feature extraction method is unsupervised and does not include spatial filtering, thus introduces no feature reduction. 

Basically, the Riemannian geometry operates on smoothly curved spaces which include the space of positive definite covariance matrices. 
In our application, spatial covariance matrices are estimated from the EEG samples and considered as points on a smoothly curved space.  
Distances between two points, or covariance matrices, are calculated on the tangent space depending on a reference point $\mathbf{C}_{ref}$.  
In contrast with the Euclidean distance which determines the shortest distance along direct paths, the Riemannian distance searches the shortest path along geodesics on curved spaces~\cite{Riemann_mean2005}. 
A dedicated Riemannian kernel~\cite{Barachant2013} calculates Riemannian features by taking the Riemannian distance into account and thus estimating the distance between covariance matrices much more accurately.

\begin{figure*}
\includegraphics[width = 1\textwidth]{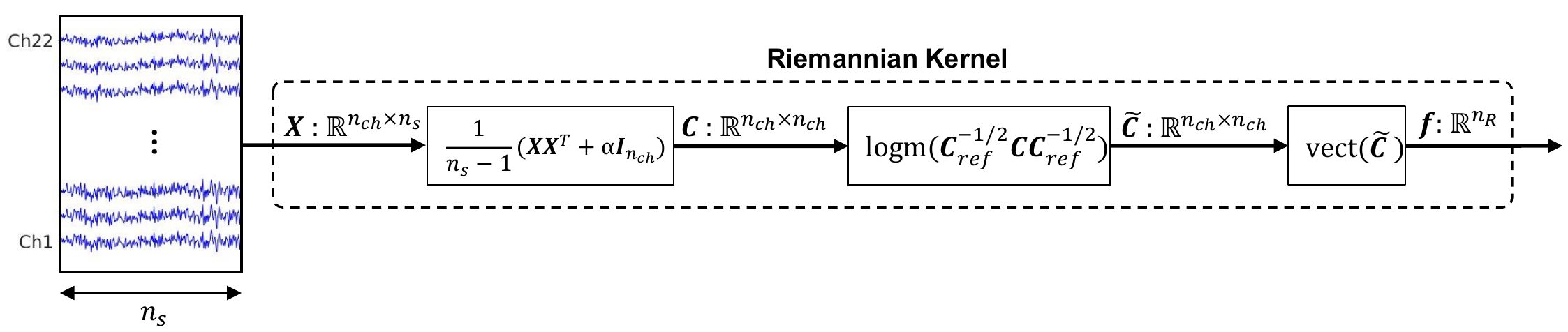}
\caption{Riemannian kernel for calculating spatial energy features of raw EEG samples.}
\label{fig:riemannian_kernel}
\end{figure*}

Fig. \ref{fig:riemannian_kernel} illustrates the computing steps to calculate Riemannian features based on the dedicated Riemannian kernel. 
The multi-channel EEG signal is extracted from a temporal window yielding a two dimensional matrix 
\begin{align}
\mathbf{X} \in \Reals ^{n_{ch} \times n_s}, 
\end{align}
where $n_{ch}$ is the number of channels and $n_s$ the number of samples of a temporal window. 
The sample covariance matrix is estimated as 
\begin{align}
\boldsymbol{C} = \frac{1}{n_s-1} (\boldsymbol{X} \boldsymbol{X}^T + \alpha \boldsymbol{I}_{n_{ch}}), 
\end{align}
where $\boldsymbol{I}_{n_{ch}}$ is the $n_{ch} \times n_{ch}$ identity matrix and $\alpha$ a regularization constant ensuring positive definiteness of the estimated covariance matrices set to 0.1.

The Riemannian kernel $K$ calculates $n_R = n_{ch}(n_{ch}+1)/2$ output features based on the input covariance matrix $\mathbf{C}$: 
\begin{align}
K: \Reals^{n_{ch} \times n_{ch}} \rightarrow \Reals^{n_R}, 
\end{align}
and is defined as
\begin{align}
\boldsymbol{f} = \textrm{vect}\left(\textrm{logm}\left(\boldsymbol{C}_{ref}^{-1/2}\boldsymbol{C}\boldsymbol{C}_{ref}^{-1/2} \right) \right), \label{eq:riemann_feat}
\end{align}
where logm(.) is the matrix logarithm and vect(.) the $\ell_2$-norm preserving half vectorization of a matrix. 
The reference covariance matrix $\mathbf{C}_{ref}$ is the average over all covariance matrices of the training set. 
The Riemannian kernel does not need labeled data and is therefore unsupervised. 
The multiplication of the covariance matrix $\mathbf{C}$ with $\mathbf{C}_{ref}^{-1/2}$ is interpreted as spatial whitening of $\mathbf{C}$.

In analogy to frequency band common spatial pattern (FBCSP), the set of Riemannian features is extended to multi-spectral features by using multiple Riemannian kernels on different frequency bands of the multi-channel EEG signal. 
The signal is divided into multiple frequency bands using a filter bank. 
A separate Riemannian kernel is used with $\mathbf{C}_{ref}$ computed solely on the corresponding frequency band. 
A recent work~\cite{Hersche2018} with high classification accuracy, suggests to use 43 overlapping frequency bands within the 4--40\,Hz band with bandwidths varying between 2--32\,Hz.

Finally, every feature is standardized to zero-mean and unit variance per feature. 
The statistics are calculated on the training set exclusively in order to keep the classification causal.
\subsection{HD Computing}
This section provides a background on HD computing.
The brain’s circuits are massive in terms of numbers of neurons and synapses, suggesting that large circuits are fundamental to the brain’s function. 
HD computing explores this idea by looking at computing with hypervectors as ultrawide words. 
It is rooted in the observation that key aspects of human memory, perception, and cognition can be explained by the mathematical properties of HD spaces, and that a powerful system of computing can be built on the rich algebra of hypervectors.
The difference between traditional computing and HD computing is apparent in the elements that we compute with.
In traditional computing, the elements are Booleans, numbers, and memory pointers, whereas in HD computing they are hypervectors. 
Hypervectors are \textit{d}-dimensional (the number of dimensions is in the thousands) and (pseudo)random with independent and identically distributed (i.i.d.) components.
They thus conform to a holographic or holistic representation: the encoded information is distributed equally over all the \textit{d} components such that no component is more responsible to store any piece of information than another. 
Such representation maximizes robustness for the most efficient use of redundancy~\cite{KanervaPentti2009HCAI}. 
Other examples of such computing structures include holographic reduced representations~\cite{PlateT.A.1995Hrr}, semantic pointer architecture~\cite{EliasmithChris2013HtBa}, binary spatter codes~\cite{Kanerva1996}, multiply--add--permute (MAP) coding~\cite{Gayler1998}, random indexing~\cite{KanervaPentii2000RIoT}, and vector symbolic architectures (VSAs)~\cite{GaylerRossW.2004VSAa}, with a quick summary in~\cite{Rahimi2017_nanoscale}. 

The number of different, nearly orthogonal hypervectors is very large when the dimensionality is in the thousands~\cite{KanervaPentti2009HCAI, Kanerva1988}. 
Two such hypervectors can now be combined into a new hypervector using simple vector-space operations, while preserving the information of the composing hypervectors with high probability. 
Computing with hypervectors begins with selecting a set of random hypervectors to represent basic objects. 
These hypervectors are also thought of as random labels. 
For example, in a language recognition application~\cite{Aditya,Rasti2016}, the letters of the alphabet as the inputs can be the basic objects, and they are assigned to random labels. 
In the same vein, in a biosignal processing application, each input electrode is assigned to a random label, independently of all the other labels. 
They serve as seed hypervectors, and they are used to make representations for more complex objects.
To generate seed hypervectors, we use binary dense codes of equally probable 0s and 1s, i.e., $\lbrace 0, 1\rbrace^d$ where $d$ is the dimensionality which is essentially a hyperparameter that can be tuned~\cite{Frady2018_seq}. 
In the following, we describe similarity measure and arithmetic operations using this code.

\subsubsection{Similarity Measurement of Hypervectors}
An essential operation in HD computing is the computation of the distance (or similarity) between two hypervectors. 
For binary hypervectors, we use the normalized Hamming distance as distance metric between two hypervectors by counting the number of unequal bits and normalize the sum by the dimensionality \textit{d}: 
\begin{align}
    \mathrm{ham}(A,B) = \frac{1}{d}\sum_{i=1}^d 1_{A(i)\neq B(i)}.
\end{align}
It can be regarded as a measure of orientation: two hypervectors with the same orientation have a normalized Hamming distance of 0, two orthogonal hypervectors have a distance of 0.5, and two hypervectors diametrically opposed have a distance of 1.
When we draw two points at random from the 10,000-$d$ space, the distance between them is close to 0.5 with high probability~\cite{Kanerva1996}. 
Hence, most points in the HD space are dissimilar or \textit{quasi-orthogonal}. 

\subsubsection{Arithmetic Operations on Hypervectors}
HD computing builds upon a well-defined set of arithmetic operations with random hypervectors.
These arithmetic operations are used for encoding and decoding patterns. 
The power and versatility of the arithmetic derives from the fact that the basic operations, namely addition and multiplication, form an algebraic structure resembling a field, to which permutations give further expressive power.

We use a variant of the MAP coding described in~\cite{Gayler1998}.
The MAP operations on hypervectors are defined as follows. 
Pointwise multiplication of two hypervectors A and B is denoted by $A \oplus B$ and corresponds to the pointwise XOR. 
Multiplication takes two hypervectors and yields a third, $A \oplus B$, that is dissimilar (approximately orthogonal) to the two and is suited for variable binding; and addition, or bundling, takes several hypervectors and yields hypervector $[A + B +...+X]$ that is maximally similar to them and is suited for representing sets. 
The brackets $[...]$ mean that the sum hypervector is clipped to $\lbrace0, 1\rbrace^d$ based on the threshold which is set to half of the number of summed elements. 
If the number is even, ties are broken by adding a random hypervector. 
Finally, the third operation is permutation, $\rho$, that rotates the coordinates of the hypervector.
A simple way to implement this is as a cyclic right shift by one position. 
All these operations have a complexity of $\mathcal{O}(d)$ and produce a $d$-dimensional hypervector.

The usefulness of HD computing comes from the nature of the operations. 
Specifically, addition produces a hypervector that is similar to the argument hypervectors---the inputs---whereas multiplication and random permutation produce a dissimilar hypervector; multiplication and permutation are invertible, addition is approximately invertible; multiplication distributes over addition; permutation distributes over both multiplication and addition; multiplication and permutation preserve similarity, meaning that two similar hypervectors are mapped to equally similar hypervectors elsewhere in the space.

Operations on hypervectors can produce results that are approximate or “noisy” and need to be associated with the “exact” hypervectors.
For that, a list of known (noise-free) seed hypervectors is maintained in a so-called “item” or “cleanup” memory. 
When presented with a noisy hypervector, the item memory (IM) outputs the hypervector that is most similar or closest. 
Making this work reliably requires high dimensionality. 
With 10,000-$d$ hypervectors, 1/3 of the bits can be flipped at random and the resulting hypervector can still be identified with the originally stored one with very high probability.

The operations make it possible to encode and manipulate sets, sequences, and lists-in essence, any data structure. 
A data record consists of a set of fields (keys, variables, or attributes) and their values (fillers).
A data record consisting of fields \textit{x, y, z} with values \textit{a, b, c} can be encoded into an hypervector H as follows. 
First, random seed hypervectors are chosen for the fields and the values \textit{(X, Y, Z, A, B, C)}, and are stored in the IM.
We then encode the record by binding the fields to their values with multiplication and by adding together the bound pairs
\begin{align}
    H = [(X \oplus A) + (Y \oplus B) + (Z \oplus C)] \label{eq:holo_eq}
\end{align}
This representation is holographic because the fields are superposed over each other---there are no spatially identifiable fields. 
Importantly, the value of \textit{x} can be extracted from this holographic representation by multiplying \textit{H} with the inverse of \textit{X}, which for $\oplus$ is \textit{X} itself:
\begin{align}
    A' = X \oplus H.  
\end{align}
The resulting hypervector \textit{A'} is given to the IM which returns \textit{A} as the most similar stored hypervector. 
%


\subsubsection{Learning and Classification with HD Computing}
\begin{figure*}
\includegraphics[width = 1\textwidth]{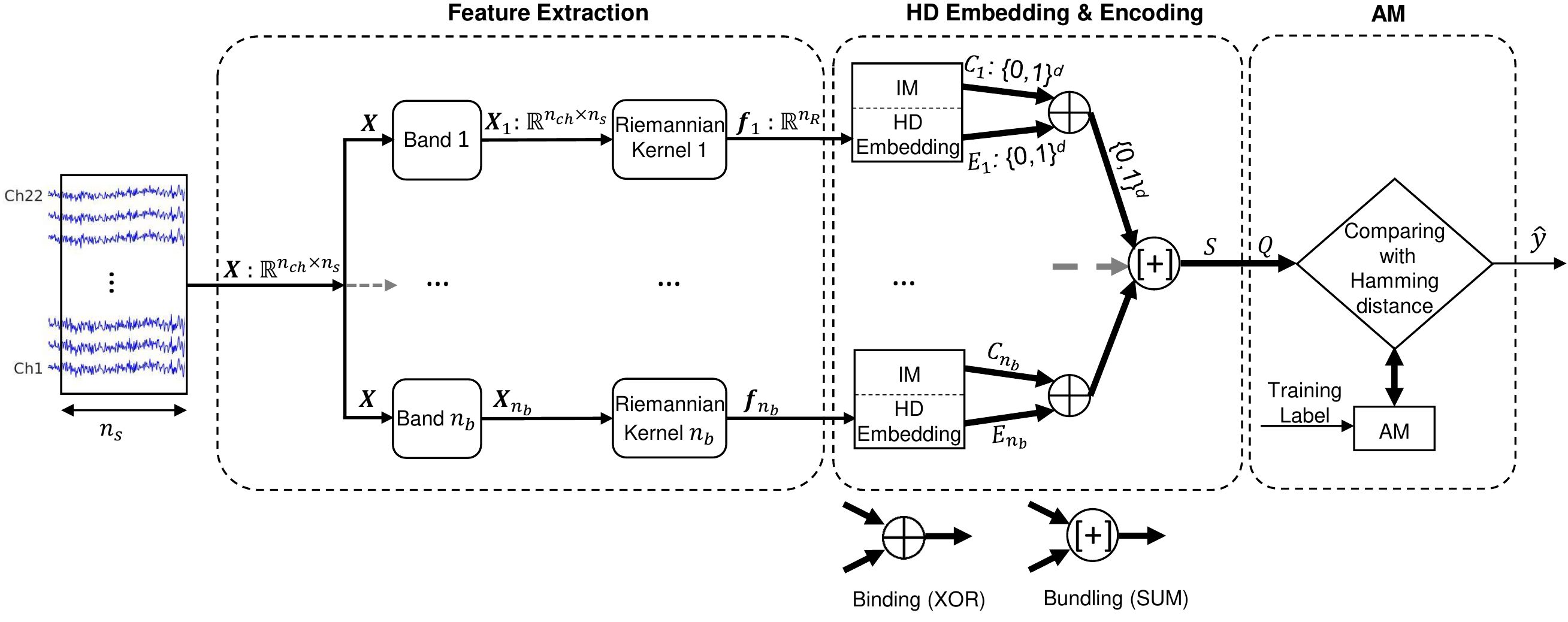}
\caption{Overall architecture for learning and classifying EEG signals with HD computing that consists of Feature Extraction, HD Embedding \& Encoding, and Associative Memory (AM). The EEG signal $\boldsymbol{X}$ of one temporal window with $n_s$ samples and $n_{ch}$ channels is processed at the time. Every EEG channel is divided into $n_b$ frequency bands ($b_1 - b_{n_b}$) using the second order Butterworth band pass filters. A Riemannian covariance kernel computes spatial energy features which are mapped to a binary hypervector using the HD Embedding \& Encoding block. The encoded hypervector is learned or classified via the associative memory (AM). During classification, AM returns the estimated class $\hat{y}$.}
\label{fig:architecture}
\end{figure*}

HD computing for learning and inference tasks is composed of four main steps: 1) embedding inputs to seed hypervectors; 2) combining seed hypervectors with the arithmetic operations inside an \emph{encoder} to produce a composite hypervector representing an object of interest; 3) combining the composite hypervectors from the same category of objects to produce a \emph{prototype} hypervector representing the entire class of category (i.e., learning); 4) finally comparing the prototype hypervectors with a query hypervector to put it into categories (i.e., inference).
These main steps are reflected in our architecture for classifying EEG signal with HD computing shown in Fig. \ref{fig:architecture}. 
Initially, the feature extraction is an extension of Fig. \ref{fig:riemannian_kernel} and calculates spectral and spatial energies out of raw time samples using a filter bank with $n_b$ frequency bands and multiple Riemannian kernels.
The output comprises $n_b$ real-valued vectors of dimension $n_R$, which is the number of features per Riemannian kernel. 
Each feature vector is mapped separately to a binary hypervector that is bound with its corresponding frequency band hypervector stored in the IM. 
The bound hypervectors are bundled to a single composite hypervector using the thresholded sum (Section \ref{sec:hd_mapping}). 
In training, the associative memory (AM) (Section \ref{sec:AM}) combines the composite hypervectors from the same MI task via thresholded bundling to one prototype hypervector per MI task. 
During inference, a new query hypervector is classified according to the most similar or closest prototype in the Hamming distance sense.  

\section{HD Embedding}\label{sec:hd_mapping}
In this section, we present the main contribution of the paper by proposing methods to represent EEG signals with binary hypervectors. 
The binary HD representation of biosignals becomes intuitive when there is a symbolization method that can map raw inputs to discrete symbols, or when the
time-domain features can be easily extracted. 
%
%
For instance, raw ECoG signal is symbolized directly with local binary patterns, and then encoded to hypervectors~\cite{BioCAS18}. 
In another application, time-domain features of EMG signals are extracted directly from the raw data and mapped to hypervectors~\cite{RahimiAbbas2016HbpA,ISCAS18}.
The MI-EEG signals however have an lower SNR than ECoG and EMG, and are usually described by an arbitrary number of mixed time-frequency features. 
The representations of such complex real-valued features in the HD space requires further investigations provided in this section.

In our encoder shown in Fig.~\ref{fig:architecture}, the HD embedding transforms a real-valued Riemannian feature vector of dimension $n_R$ to a $d$-dimensional hypervectors. 
The hypervectors of every band are bound with its corresponding frequency band hypervector stored in the IM.
The bound hypervectors are bundled together to form a complex representation of the multi-spectral signal. 
The same HD embedding function is shared among all frequency bands.

The encoder is inspired by the holographic data record structure in Eq.~\eqref{eq:holo_eq}, where here the values are defined as $E_1 \ldots E_{n_b}$ with their corresponding fields as $C_1 \ldots C_{n_b}$.
At initialization, the fields $C_1 \ldots C_{n_b}$, or frequency band hypervectors, are drawn at random and stored in the IM. 
In the example of Eq.~\eqref{eq:holo_eq} the values are already stored in the IM, however, here they are the output of an HD embedding method. 
An embedding is a representation for which the computation of distances directly gives an estimate of the distances in their initial representation~\cite{Rachkovskij2017}. 
The building of such representations is provided by binary locality-sensitive hashing (LSH) functions,  which ensure that similar elements are statistically likely to be embedded into the same value~\cite{LSH_EMBC15}. 
Once mapped to the HD space, the similarity is computed with the Hamming distance. 

In order to produce hypervectors in the HD sense we aim to fulfill the following constraints: 
\begin{itemize}
\item\textbf{Holographic representation}\\
The performance of HD computing degrades significantly if components of the hypervector are correlated~\cite{KellyMatthewA2013}.
Hence, the encoder aims to produce holographic or distributed hypervectors with no localized information. 
If the embedding does not generate distributed hypervectors, subsequent (pseudo)random shuffling or binding with a random hypervector establishes distributed representation.

\item\textbf{Zero-mean output}\\
The hypervectors have to be zero-mean which corresponds to an equal number of 1s and 0s in the binary case.  
If the embedding fails to produce a zero-mean output, binding with a random hypervector taken from the IM establishes the zero-mean distribution.
\item\textbf{Similarity-preserving}\\
As opposed to the two previous constraints, similarity-preserving is not a requirement for HD representation per se.
However, the encoding may aim similarities in the real valued feature space to be preserved in the embedded HD space.
%
\end{itemize}

In the following, we propose three embedding methods encode real-valued feature vectors to the binary HD space ($\lbrace 0,1 \rbrace ^d$). 

\subsection{Individual Feature Quantization}
The simplest embedding quantizes every component of the feature vector to $l$ levels and encodes them separately using $q$ bits per value. 
All binarized values are concatenated to form a vector of dimension  $d = n_R \cdot q$.
This representation is inherently not distributed since the information of every vector entry can be localized precisely. 
However, the use of a random permutation (not just cyclic shift) would ensure distributed representation. 
Before the quantization, every feature vector is standardized to zero-mean and unit variance separately. 
The statistics are calculated on the current vector exclusively.

We propose the use of two different codes, namely the thermometer (or unary) code and a variation of the Gray code.
In thermometer coding, the quantization level $i \in \lbrace 0,..., l-1 \rbrace$ is represented by $i$ ones followed by $(l-i)$ zeros.
The number of quantization levels is equal to the number of encoded bits per component, i.e. $q = l$. 

A simple variation of the Gray code changes two bits between two adjacent quantization levels instead of one bit.
The first level is represented by the all-zero string. 
This code has $l = \binom{q}{2}$ quantization levels and therefore has generally higher resolution than the thermometer code when keeping $q$ fixed. 
In contrast to the thermometer code, the Gray code  appears more randomized. 
However, macroscopic information about a feature vector entry can still be localized and random permutation is necessary to ensure a distributed representation

\subsection{Random Projections}
Random projections are usually used for dimensionality reduction in the Euclidean space~\cite{Bingham2001}. 
The Johnson-Lindenstrauss lemma~\cite{JohnsonLindenstrauss1982} ensures distances between two points in the projected space to be preserved if the output dimension is suitably high. 
Such projections deal with embeddings between Euclidean spaces.
However, here the data is embedded to a high-dimensional Hamming space.
Recently, it has been shown~\cite{Rachkovskij2017} that random projections can indeed project data to a high-dimensional Hamming space while preserving the distance between points with success full application in monitoring arterial blood pressure via electrocardiography (ECG) signals~\cite{LSH_EMBC15,LSH_EMBC16}.

The random projection to the high-dimensional space is defined as 
\begin{align}
   E = \textrm{H}(\boldsymbol{R} \boldsymbol{f}), 
\end{align}
where H(.) is the component-wise Heavyside step function 
\begin{align}
 \textrm{H}(z)_i  = 
   \begin{cases}
    1  & \textrm{if } z_i \geq 0 \\
    0  & \textrm{if } z_i < 0,   \\
   \end{cases}
\end{align}
and $\boldsymbol{R} \in \Reals ^{d \times n_R}$ the projection matrix \cite{Rachkovskij2017}.
Usually, the components $r_{i,j}$ of $\boldsymbol{R}$ are drawn from an i.i.d. Gaussian normal distribution ($r_{i,j} \sim \mathcal{N}(0,1)$). 
However, the Gaussian projection matrix can be replaced by a much simpler one such as the sparse bipolar random matrix \cite{Achlioptas2001}:
\begin{align}
 r_{i,j} = 
   \begin{cases}
    +1  & \textrm{with probability } s/2 \\
    0  & \textrm{with probability } 1-s \\
    -1  & \textrm{with probability } s/2, \\
   \end{cases}
\end{align}
where $s \in [0,1]$ is the relative sparsity.
Achlioptas \cite{Achlioptas2001} has shown that by using a sparsity of $s=2/3$ this projection comes without any sacrifice in the quality of embedding compared to the plain Gaussian projection. 
Moreover, this distribution reduces the computational cost, since the projection requires only additions and subtractions. 
The sparsity $s$ allows to further reduce the number of operations as many entries of the projection matrix are zero.
Furthermore, the projection matrix does not need to be restored from a memory as it can be cheaply recomputed. 
The same projection matrix is also shared among all the frequency bands to reduce the memory footprint during a full parallel projection. 
In our case we could reduce the sparsity to $s=1/10$ without loosing performance in classification accuracy.

Random projections can be viewed as a collection of $d$ randomly drawn hyperplanes in the $n_R$-dimensional space. 
A new point is transformed to the hyperspace by determining on which side of the hyperplane the point lies. 
The resulting HD representation is both distributed and zero-mean.
Every binary entry in the hypervector depends on the linear combination of most of the input features. 
Therefore, information about a specific input feature component cannot be localized in the projected hypervector.

\subsection{End-to-end Learned Projections}
Instead of randomly projecting data to the HD space, projections can be learned to form a more discriminative representation. 
Well-known examples of learning projections are PCA or singular value decomposition (SVD)~\cite{Keogh2001}.
However, they are used for dimensionality reduction in the Euclidian space. 
This section presents a novel embedding strategy, which learns to project a real valued feature space to the binary HD space.
In principle, the embedding learns to maximize distances between hypervectors originating from input vectors of different classes. 
In the input space, two vectors of different classes are not necessarily far apart, depending on the quality of the feature extraction. 
Nevertheless, the aim is to ``tear" those two vectors apart in the HD space in order to achieve high classification accuracy. %
This embedding does not fulfill the similarity-preserving condition.
Instead, it increases the performance of the overall algorithm. 

\begin{figure}
    \centering
    \includegraphics[width=0.45\textwidth]{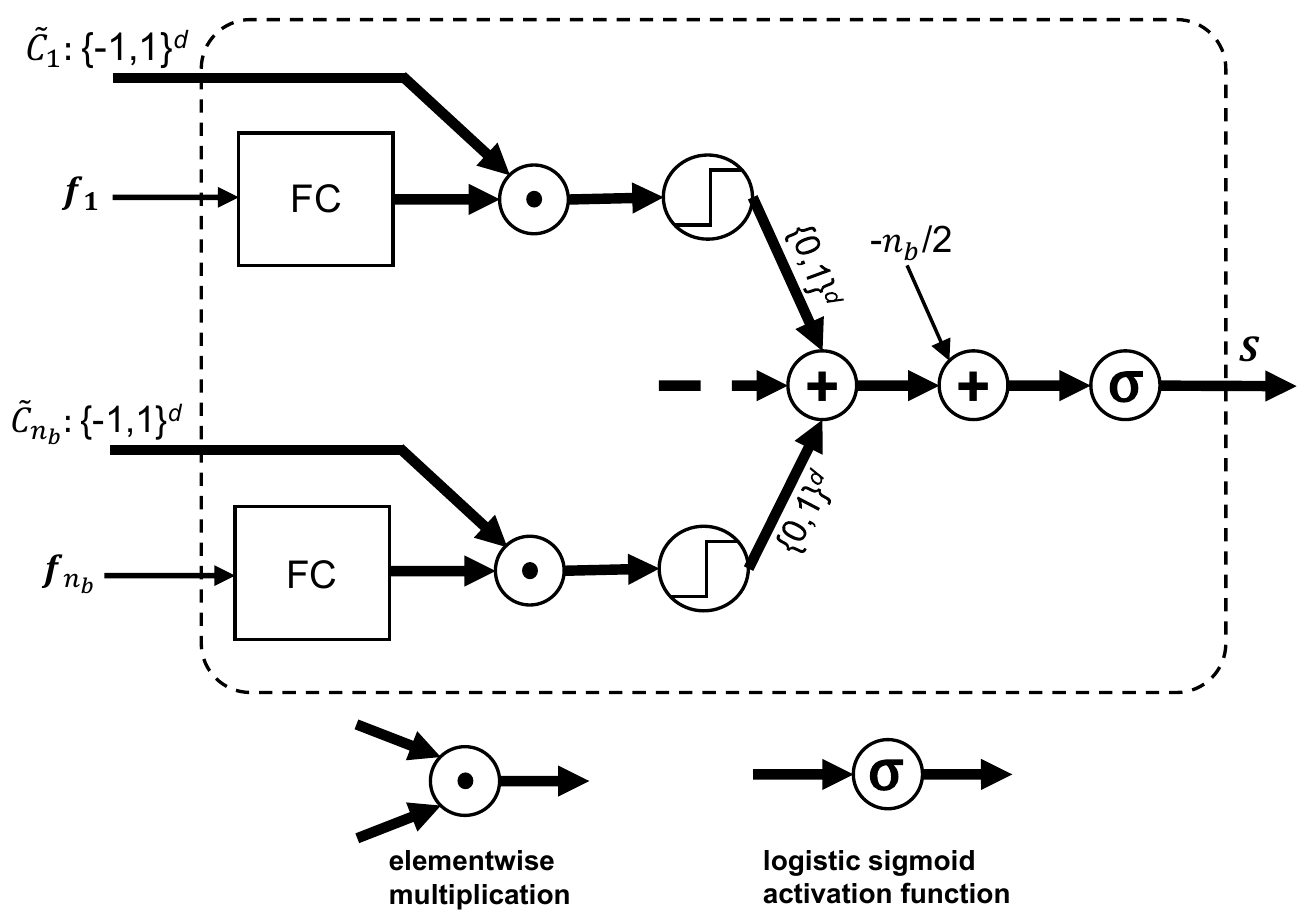}
    \caption{Transformed encoder architecture for training the HD embedding on its binary neural net (BNN) equivalence. All vectors are implemented in floating point arithmetic, even though they may have a limited number of values, e.g. $\lbrace-1,1\rbrace$. The HD embedding and binding is replaced by a fully connected (FC) linear layer without bias, an component-wise multiplication with the corresponding transformed frequency band hypervector $\tilde{C}$ defined in \eqref{eq:ctilde} and a discretization component. } 
    \label{fig:end_end_learning}
\end{figure}

Our method learns the projection end-to-end on the encoder using backpropagation.
In training, the HD embedding \& encoder block in Fig.~\ref{fig:architecture} is transformed to its neural net equivalence shown in Fig.~\ref{fig:end_end_learning}.
The network simulates a binary HD network. 
In training, however, it is implemented with real valued logits and gradients. 
The projection corresponds to a fully connected (FC) layer without a bias, where the weights of the projection matrix $\mathbf{R}$ are shared among all frequency bands.
In contrast to the sparse bipolar random projection matrix the FC weights are dense and real-valued. 
The component-wise XOR operation is placed before the quantization. 
The equivalent operation before the quantization is the component wise multiplication with the transformed frequency band hypervectors
\begin{align}
\tilde{C}_i = -2 C_i + 1 \quad \quad i \in \lbrace 1,2,...,n_b \rbrace. 
\label{eq:ctilde}
\end{align}
The thresholding after the summation of all bands in the encoder is replaced by a logistic sigmoid function for simpler training. 

At the beginning of the training, one target vector $P^*$ per class is drawn at random.
Therefore, the class hypervectors are quasi-orthogonal to each other, irrespective of the relations of the input feature vectors. 
The goal is to minimize the binary cross entropy (BCE) loss between the output $S$ and the target vector $P^*$. 
The optimization problem is derived from multi-class logistic regression where each class is encoded to a one-hot vector of dimension $n_{cl}$, e.g. the class two would be encoded to (0,1,0,0) when having $n_{cl}=4$ classes.
In our case the encoded target vector $P^*$ is a hypervector of dimension $d>>n_{cl}$, however, the same optimization problem can be applied. 

Backpropagation together with stochastic gradient descent (SGD) minimizes the BCE loss function.
The derivative of the Heaviside function after the FC is zero almost everywhere, which makes it difficult to propagate the gradient through the discretization layer.
Hence, the following strategy commonly used in binary neural networks (BNN)~\cite{MontoneGuglielmo2017Hcfa} is applied.
Given the Heaviside discretization function 
\begin{align}
    q = \textrm{H}(r), 
\end{align}
and the cost function $L$, the estimation $g_q$ of the gradient $\frac{\partial L}{\partial r}$ is propagated using the ``straight-through estimator" 
\begin{align}
    g_r = g_q 1_{|r|\leq 1 }. 
\end{align}
The gradient is simply passed through the discretization if the absolute value of the input is below or equal to one. 

\section{Associative Memory}\label{sec:AM}
\begin{figure}
    \centering
    \includegraphics[width=0.45\textwidth]{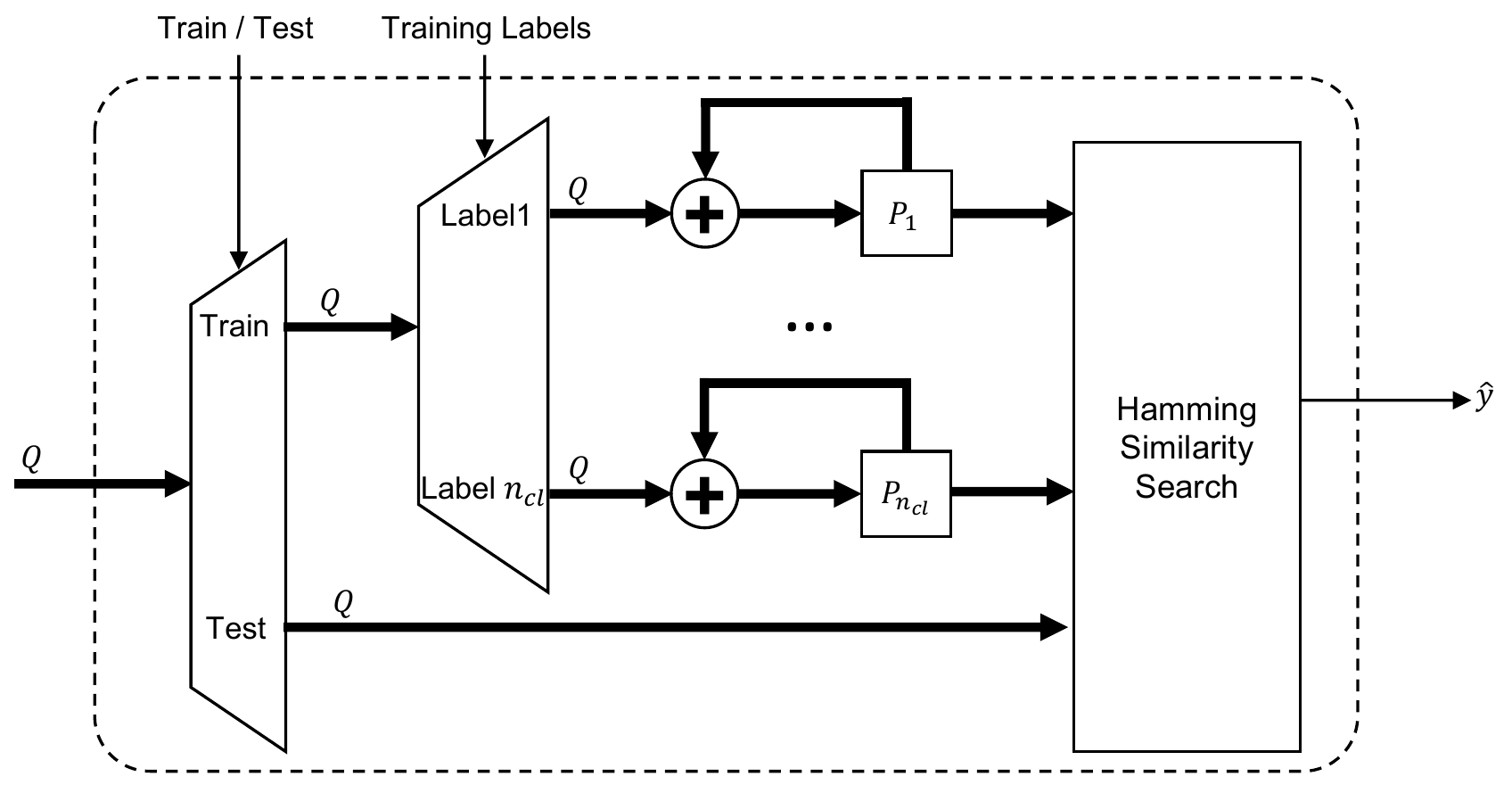}
    \caption{Associative memory (AM) classifier. } 
    \label{fig:am_learning}
\end{figure}

The previous section describes a set of embedding methods that map feature vectors to a binary hypervector.
This section focuses on associative memory (AM) that completes the supervised learning procedure by assigning a label to the output hypervector of embedding method.
Fig.~\ref{fig:am_learning} illustrates the functionality of the AM . 
Basically, the AM supports a training and a testing mode. 
In training, the AM aims to learn one prototype hypervector ($P$) per class. 
The class prototype is a hypervector representing all items from the entire class. 
The AM adds an encoded hypervector $Q$ to the corresponding prototype ($P_i$) based on provided label of $i$ 
\begin{align}
    P_i += Q \quad  i \in \lbrace 1,2,...,n_{cl}\rbrace, 
\end{align}
where $n_{cl}$ corresponds to the number of classes. 
Even though the output of the encoder ($Q$) is a binary vector, it is treated as an unsigned integer vector in training.
The number of additions $n_i$ is tracked for every prototype separately and is used for the threshold calculation.
In the end of the training, the accumulated prototype vectors are binarized: 
\begin{align}
    P_i \xleftarrow[]{} \textrm{H}(P_i-n_i/2). 
\end{align}
This prototype generation can be interpreted as a sort of \textit{averaging}. 
If the number of additions $n_i$ is even, ties are broken at random by adding a random hypervector. 

When using the spatial encoder, it is beneficial to skip the thresholding after the additions in order to increase the resolution of the AM learning. 
The entries of the embedding hypervector have then values in $\{0,1,...,n_b\}$ instead of binary values $\{0,1\}$. 
The only change in the AM learning is that the number of additions $n_i$ is increased by factor $n_b$ with every new hypervector. 
%

\begin{figure}
    \centering
  \subfloat[Learned projection]{\label{fig:sub1}{\includegraphics[width=0.2\textwidth]{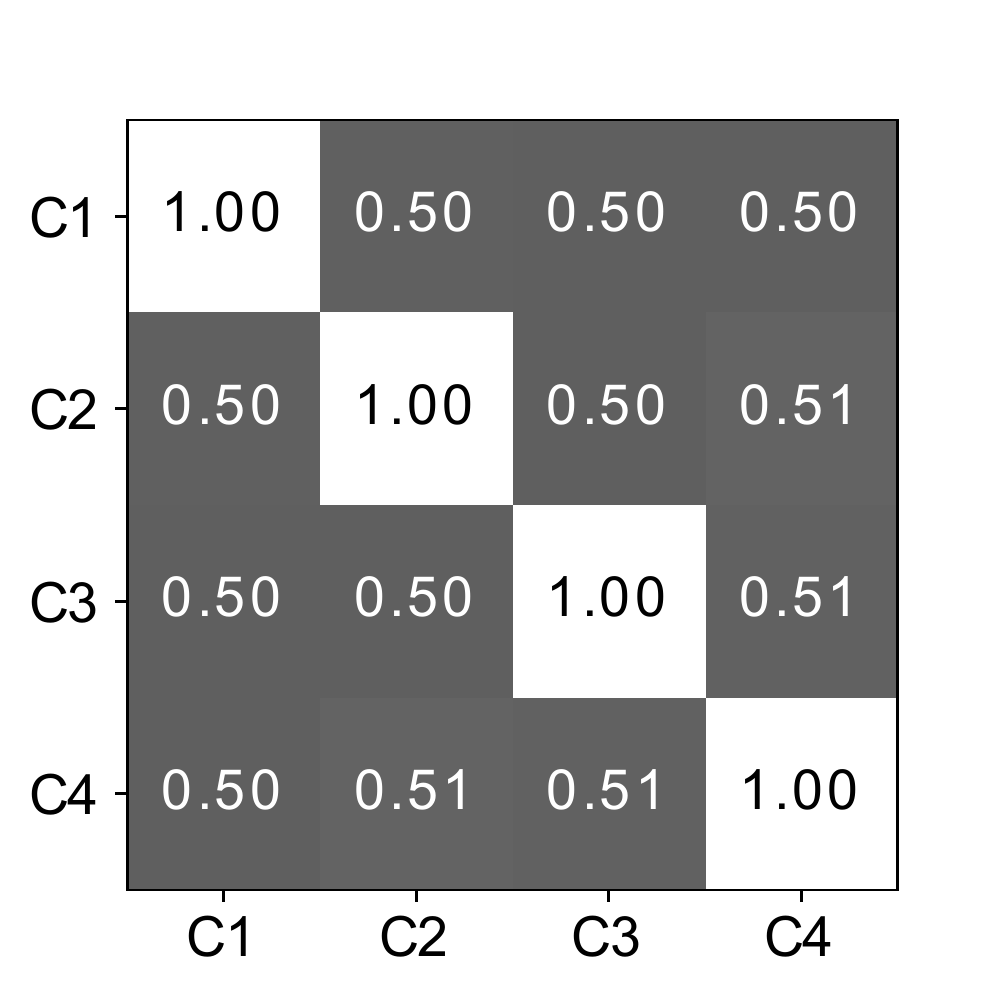}}}
\subfloat[Random projection]{\label{fig:sub2}{\includegraphics[width=0.25\textwidth]{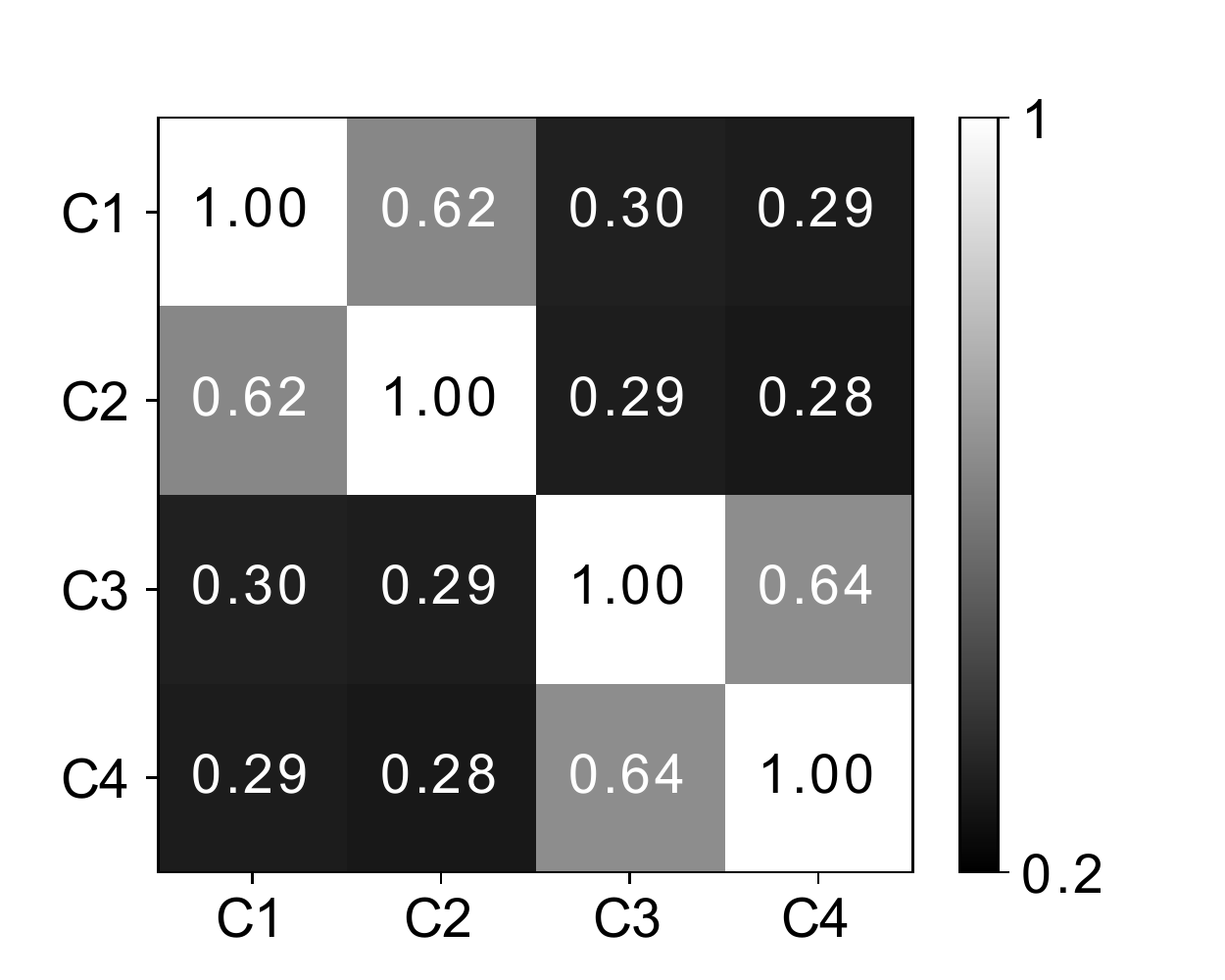}}}

\caption{Relative similarities between the prototypes in the associative memory (AM) with (a) learned projections and (b) random projections, both with $d=10,000$. Learned projection is capable to produce quasi-orthogonal prototype hypervectors which are highly dissimilar. Random projection however yields prototype hypervectors which are much more similar, especially between C1--C2 and C3--C4.}
\label{fig:similarities}
\end{figure}

If the architecture uses learned projections as HD embedding, there is no need to train the AM memory again. 
The prototype hypervectors are simply the target hypervectors assigned to every class used in the projection training.
Fig.~\ref{fig:similarities} illustrates the difference in the AM when generating the AM by averaging or by end-to-end HD training. 
The AM prototypes trained by averaging are much more similar to each other compared with those generated by end-to-end HD training. 
%
%

During inference, the encoded query hypervector ($Q$) is classified by searching for the AM prototype with the shortest relative Hamming distance.
\begin{align}
    \hat{y} = \operatornamewithlimits{\textrm{argmin}}_{i=1,...,n_{cl}} \left( \textrm{ham}(P_i,Q) \right)
\end{align}

\subsection{Associative Memory with Multiple Prototypes}\label{subsec:AM_mult_prot}
We extend the AM by learning multiple prototypes per class using a k-means clustering algorithm with HD operations called as HD k-means.
The clustering is done for every class separately. 
Basically, the HD k-means clustering differs from the Euclidean k-means only in the distance measure and the mean calculation.
It uses the Hamming distance and the thresholded bundling explained in the previous section.
The clustering used in this paper is an extension called k-means++ which runs the k-means multiple times with different initial centroids chosen randomly from the training set.

After the clustering, the AM contains k prototypes per class, which are the cluster centers of the HD k-means algorithm. 
This AM generation obviously requires more computational power than simple AM average learning due to the multiple iterations the k-means++ clustering has to do.  
In inference, the Hamming distance between the query hypervector $Q$ and all the prototypes is calculated. 
The estimated class is the one which contains the cluster center with the minimum distance to the query hypervector:
\begin{align}
    \hat{y} = \operatornamewithlimits{\textrm{argmin}}_{i=1,...,n_{cl}} \left(\operatornamewithlimits{\textrm{min}}_{a=1,...,k} \textrm{ham}(P_i^a,Q) \right)
\end{align}

\section{Experimental Results}\label{sec:results}
In this section, we assess the proposed methods on two different datasets containing EEG data of MI experiments. 
We measure classification accuracy, memory footprint of the model, execution time and energy consumption during training and inference.
The classification accuracy is defined as
\begin{align}
\textrm{classification accuracy} = \left( \frac{n_{\textrm{correct}}}{n_{\textrm{total}}} \right) \times 100 \%, 
\end{align} 
where $n_{\textrm{correct}}$ is the number of correct classified trials and $n_{\textrm{total}}$ the total number of trials in the test set per subject.
Apart from the classification accuracy, the classifiers are compared in terms of their memory footprint and computational complexity.  
The memory footprint is straight-forward to analyze, however, an equal comparison of the computational complexity between floating point operations and binary operations is not trivial. 
Therefore, the run time and energy consumption of training and testing is measured on an NVIDIA TX2.
Nvidia's Tegra X2 platform is an embedded computing SoC targeted at AI workloads. 
It consists of a dual-core Denver2 ARMv8 CPU, a 4-core ARM Cortex-A57 (both running at 2\,GHz), and a 256-core GPU based on the Pascal architecture running at 1.46\,GHz, which provides a performance of 750\,GFLOPS (single-precision) at a power consumption of around 15\,W and has a memory bandwidth of 58.4\,GB/s. 
For comparison, this is corresponds to 1/14 the compute units (at same speed) and 12\% of the memory bandwidth of the recent GTX~1080~Ti desktop GPU. 
The TX2 has several power modes, which allows to perform dynamic voltage-frequency scaling. 
In the Max-N mode all the processing units work at maximum frequency, whereas in the Max-Q mode only the ARM cluster and the GPU are active, both running at a lower frequency (\SI{1.2}{\GHz} for CPU and \SI{0.85}{\GHz} for GPU), for maximum energy efficiency.
We use both modes for all our experiments on the Jetson TX2 development board, and the on-board sensors for power measurements.

An $\ell_2$-regularized linear SVM has performed best on the BCI Competition IV-2a dataset~\cite{BCI-recording} with multi-spectral Riemannian features~\cite{Hersche2018} and serves as baseline classifier.
The regularization constant $c$ is determined by a grid search in cross validation and is set to 0.1.
The SVM uses the same Riemannian features as the HD classifier, but they are concatenated to one vector and fed directly to the SVM~\cite{Hersche2018}. 

\subsection{Dataset Descriptions}\label{sec:datasets}

\begin{figure}[hptb]
\centering
\includegraphics[width=0.5\textwidth]{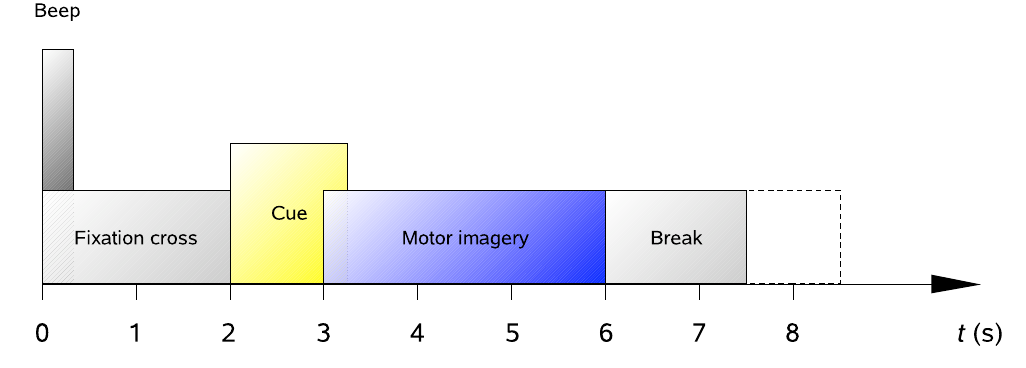}
\caption{Timing scheme of BCI competition IV2a dataset~\cite{BCI-recording}.\label{fig:timing_IV2a}}
\end{figure}

\textbf{a) 4-class MI dataset.} 
The BCI Competition IV-2a dataset~\cite{BCI-recording} consists of EEG data from 9 different subjects. 
The subjects were requested to carry out four different MI tasks, namely the imagination of the movement of the left hand, right hand, both feet and tongue. 
Two sessions were recorded on two different days. 
For each subject a session consists of 72 trials per class yielding 288 trials in total. 
One session is used for training and the other for testing exclusively. 
The signal was recorded using 22 EEG electrodes according to the 10-20 system. 
It is bandpass filtered between 0.5\,Hz and 100\,Hz and sampled with 250\,Hz. 
In addition to the 22 EEG channels, three electrooculography (EOG) channels give information about the eye movement.  
An expert marked trials containing artifacts based on the EOG signal. 
This way 9.41\% of the trials were excluded from the dataset. The number of trials per class remains approximately balanced.
The Riemannian feature extraction uses $n_b = 43$ overlapping frequency bands in the range between 4--40\,Hz with bandwidths varying within 2--32\,Hz described in~\cite{Hersche2018}.
The classification is done on a temporal window from 2.5\,s to 6\,s according to the timing scheme illustrated in Fig \ref{fig:timing_IV2a}.
When using 22 EEG channels the number of Riemannian features per frequency band is $n_R = 22(22+1)/2 = 253$.
This gives a total of $n_b \cdot n_R = 10,879$ features when using 43 frequency bands.  

\textbf{b) 3-class MI dataset.}
The second dataset is based on the study in~\cite{Saeedi2016}, and consists of 3-class motor imagery (left hand, right hand and feet) from 5 different subjects.
Every subject participated in four sessions recorded on the same day including 45 trials per session. 
The signal was recorded with 16 EEG electrodes with a sampling frequency of 512\,Hz. 
This dataset does not include any EOG information, therefore, no trials are excluded due to any artifacts. 
For testing, 4-fold cross validation is applied, using three sessions for training and one for testing.
In this dataset it turned out to be sufficient to use only $n_b = 13$ frequency bands in the range of 4--30\,Hz with bandwidth 2\,Hz and a temporal window of 4\,s length. 
The use of more frequency bands does not hurt the classification performance, however, it increases the computation time. 
The resulting number of Riemannian features per frequency band is $n_R = 16(16+1)/2 = 136$ and the total number of features $1768$.

\subsection{MI classification}\label{sec:miclassification}
\begin{table*}[htpb]
\vspace{0.05in}
\centering
\caption{Summary of the best classification accuracies (\%) on the 4-class MI BCI competition IV2a test set and the  4-fold cross-validation accuracy on a 3-class MI data set. The performance of HD classifiers with different binary HD embeddings is compared with a linear $\ell_2$-regularized SVM (c = 0.1). All classifier are fed with the same multi-spectral Riemannian features.}
\label{tab:results}
\begin{tabular}{clcccccc}
\cmidrule(r){1-8}  
& Subject &	 SVM &	 \begin{tabular}[c]{@{}c@{}}HD \\ thermometer\end{tabular} &	 HD Gray &	 \begin{tabular}[c]{@{}c@{}}HD random \\ projections\end{tabular}  &	 \begin{tabular}[c]{@{}c@{}}HD learned \\ projections\end{tabular} & \begin{tabular}[c]{@{}c@{}}HD learned \\ k=3 \end{tabular}\\ 
 \cmidrule(r){1-8}
  & A1 &	91.46 &	86.05 &	81.32 &	84.52 &	88.40 & 86.83\\ 
& A2 &	52.65 &	57.92 &	54.77 &	54.81 &	54.17 & 54.50\\ 
& A3 &	83.52 &	71.10 &	70.70 &	67.91 &	78.35 & 80.14\\ 
& A4 &	67.54 &	67.11 &	67.32 &	71.54 &	68.60 & 64.86\\ 
4-class& A5 &	63.77 &	56.96 &	57.61 &	55.33 &	61.12 & 62.47\\ 
MI$^{\textrm{(a)}}$ & A6 &	60.47 &	50.79 &	47.44 &	50.84 &	56.33 & 57.26\\ 
& A7 &	87.36 &	68.63 &	67.33 &	67.40 &	85.49 & 86.64\\ 
& A8 &	81.18 &	73.25 &	74.72 &	72.47 &	77.38 & 77.12\\ 
& A9 &	80.68 &	79.24 &	76.89 &	69.58 &	81.14 & 83.08\\ 
\cmidrule(r){2-8}  
& \textbf{{Mean}} &	 \textbf{74.29} &	 \textbf{67.89} &	 \textbf{66.46} &	 \textbf{66.04} &	 \textbf{72.33} & \textbf{72.54}\\ 
& Std &	12.06 &	10.05 &	9.97 &	9.50 &	11.38 & 12.17\\ 
\cmidrule(r){1-8}  
& B1 &	77.22 &	68.50 &	64.39 &	65.83 &	80.00 & 79.89 \\
& B2 &	82.78 &	76.06 &	71.33 &	72.44 &	81.89 & 83.61\\ 
3-class & B3 &	89.44 &	86.72 &	86.67 &	85.28 &	87.78 & 88.33\\ 
MI$^{\textrm{(b)}}$ & B4 &	98.33 &	97.89 &	94.67 &	95.78 &	98.00 & 98.33\\ 
& B5 &	65.56 &	69.28 &	70.33 &	73.44 &	69.94 & 70.94\\ 
\cmidrule(r){2-8}  
& \textbf{{Mean}} &	 \textbf{82.67} &	 \textbf{79.69} &	 \textbf{77.48} &	 \textbf{78.56} &	 \textbf{83.52} & \textbf{84.22}\\ 
& Std &	10.11 &	10.23 &	10.33 &	9.72 &	8.44 & 9.07\\ 
\cmidrule(r){1-8}  
\end{tabular}
\end{table*}
%
This section summarizes the maximum classification accuracy achieved by the individual classifier and HD embedding configurations. 
A more elaborate analysis will be given in the next sections. 
Table~\ref{tab:results} compares the classification accuracy on the 4- and 3-class datasets for the linear SVM and HD classifiers using different embedding methods.

The linear SVM achieves 74.29\% and 82.67\% average classification accuracy on the 4- and 3-class dataset, respectively. 
%
%
The HD classifier is outperformed by the linear SVM on both datasets when using simple thermometer or Gray code as well as random projections. 
Among this three HD embeddings, the thermometer code shows the highest classification accuracy of 67.89\% and 79.69\%, which is 6.40\% and 2.89\% lower compared to the linear SVM. 
The accuracy of the HD classifier improves significantly when using the learned projections instead, achieving 72.33\% and 83.52\% average classification accuracy. 
It is further improved by using multiple prototypes in the associative memory. 
The last column of the table shows that maximum achieved classification accuracy of 72.54\% and 84.22\% using learned projections with k=3 prototypes per class. 
On the 3-class dataset, the HD classifier outperforms the linear SVM by 1.55\%. 
Even though the maximum accuracy of the HD classifier is higher than the SVM on the 3-class and slightly lower on the 4-class dataset, the standard deviation of the HD accuracy is close or lower than the SVM in both datasets supporting the statistic significance of the results. 

\subsection{Dimensionality of Embedding}
This section examines the performance of different HD embbeding methods, considering not only the classification accuracy but also the necessary dimension $d$.
Even though operations in the binary HD space are relatively cheap, the dimensionality should be tuned to reduce the number of operations.

\begin{figure*}[htbp]
  \subfloat[4-class MI]{\label{fig:sub_hd_map_results_a}{\includegraphics[width=0.475\textwidth]{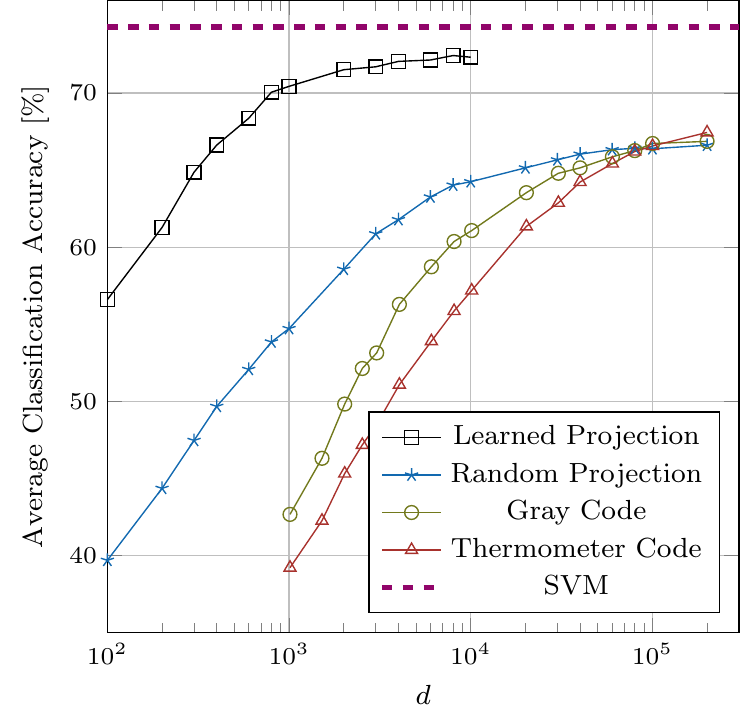}}}\hfill
\subfloat[3-class MI]{\label{fig:sub_hd_map_results_b}{\includegraphics[width=0.475\textwidth]{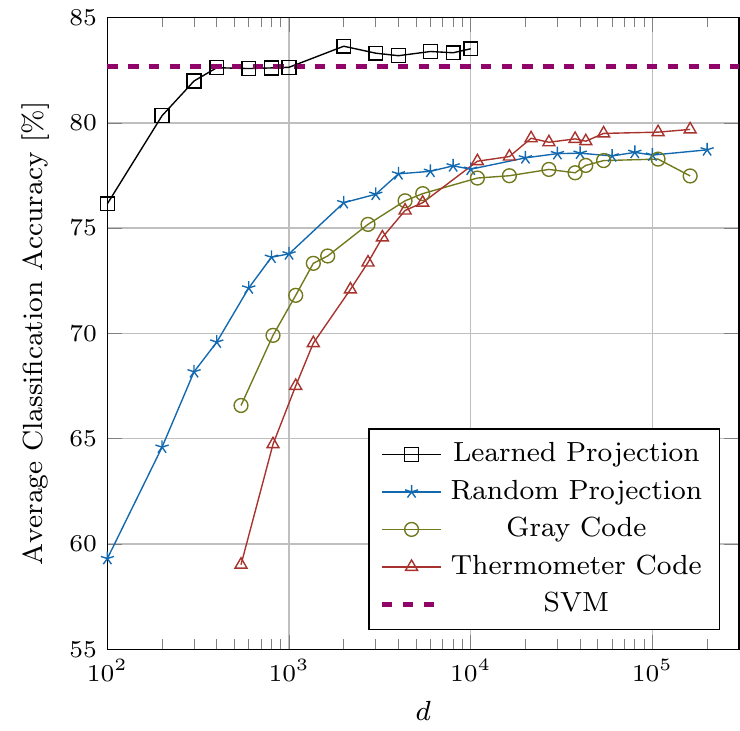}}}
\caption{Classification accuracy of different HD embedding configurations compared with linear SVM.}
\label{fig:hd_map_results}
\end{figure*}

Fig.~\ref{fig:hd_map_results} shows the classification accuracy on both datasets using different embedding methods.
The random projection matrix consists of tripolar values $\lbrace-1,0,1\rbrace$, where 90\% of the values are set to zero, corresponding to a relative sparsity of $s=0.1$.
Experiments with Gaussian random projection matrices yielded the same results as with the sparse tripolar projection, thus, it is excluded from the figure. 
In this experiments, a simple associative memory classifier with one prototype per class is used. 
Due to the randomness of the HD classifier, the results are averaged over 10 runs in order to get reliable results. 

The performance of random projections, Gray and thermometer codes is considered first. 
Among this HD embedding the thermometer code performs worst in lower dimensions. 
However, it outperforms Gray coding and random projections when going to higher dimension above $100,000$ on the 4-class dataset and $10,000$ on the 3-class dataset.
Random projections on the other hand, exploit the lower dimensions better than Gray and thermometer coding, but converge approximately to the same accuracy in high dimensions.
%

The learned projection outperforms the aforementioned embeddings in all aspects by achieving the highest classification accuracy in lower dimensions. 
On the 4-class dataset, however, it still does not outperform the linear SVM. 
Conversely, the learned projection requires only $d$=400 to perform on par with the linear SVM on the 3-class dataset.
It effectively compresses a real feature space of $13\times136$ to a binary space of 400. 
As soon as going above $d$=1,000, it outperforms SVM by a small margin of 0.85\%, and eventually by 1.55\% at $d$=10,000. 


\subsection{Associative Memory with Multiple Prototypes}
After discussing the performance of various embedding methods, the AM is improved by using multiple prototypes per class with the method proposed in Section~\ref{subsec:AM_mult_prot}. 
The AM uses an HD-version of the k-means++ clustering algorithm. %
It runs the same clustering 10 times with different initial centroids chosen randomly from all the hypervectors in the training set. 
This section discusses the results when using learned projections with fixed HD dimension $d$=8,000, which showed to be high enough on both datasets according to the previous section. 
In training, the projection is first trained to produce one prototype per class. 
In the second phase, the projection is fixed and the AM is trained with a higher number of prototypes per class. 
The results reported here are averaged over 10 runs.

\begin{figure*}[htbp]
  \subfloat[4-class MI]{\label{fig:sub_hd_map_results_a}{\includegraphics[width=0.475\textwidth]{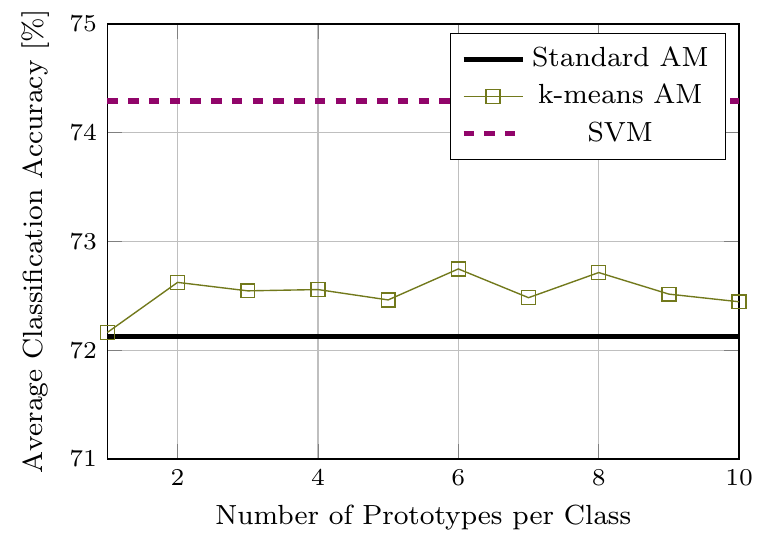}}}\hfill
\subfloat[3-class MI]{\label{fig:sub_hd_map_results_b}{\includegraphics[width=0.475\textwidth]{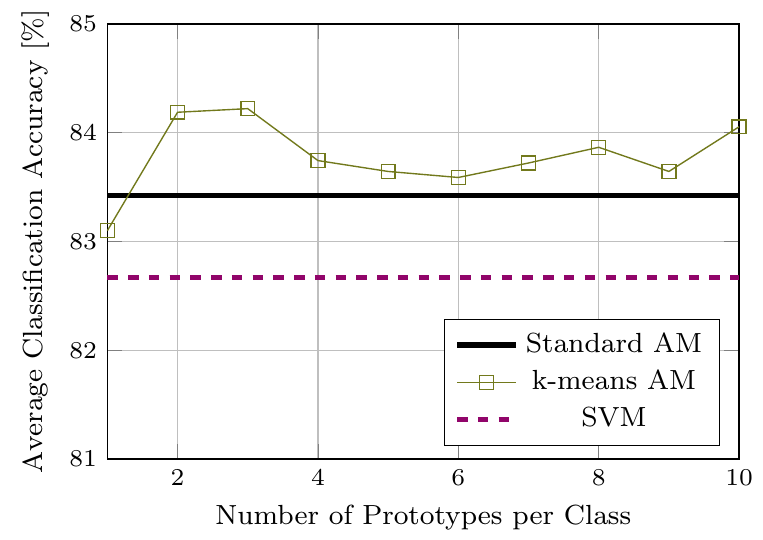}}}
\caption{Classification accuracy using k-means associative memory compared with standard associative memory and linear SVM. Learned projections with \textit{d} = 8,000 are used for HD embedding.}
\label{fig:kmeans}
\end{figure*}

Fig.~\ref{fig:kmeans} illustrates the gain on both datasets when using multiple prototypes in the AM. 
The performance of the standard AM is always surpassed when using more than one prototype per class. 
This shows the amenability of AM for incremental and online learning.
On the 4-class dataset, the maximum classification accuracy is 72.75\% achieved with $k$=6 prototypes per class. 
This is an improvement of 0.62\% compared to the standard AM. 
On the 3-class dataset, $k$=3 prototypes per class are sufficient to achieve the maximum accuracy of 84.22\%, corresponding to a gain of 0.80\%. 
When using only one prototype with the k-means AM, the accuracy is even slightly lower than the standard AM.
%
This is due to the different training procedure. 
The standard AM is trained on the unclipped versions of the hypervectors while the k-means AM trains only with clipped binary vectors.

\subsection{Memory Footprint Analysis}
This section analyses the memory footprint to store different classifier models. 
This comparison considers only the classifier, the feature extraction is neglected. 
Table~\ref{tab:memory} shows the memory footprint of the separate modules. 
The calculations are based on the following decisions: 

\begin{itemize}
    \item \textbf{SVM} \\
    The linear SVM stores $n_{cl}$ vectors of dimension ($n_R \cdot n_b$). 
    The SVM is implemented in \textit{double64}, therefore, the value is multiplied by factor 64. 
    \item \textbf{AM classification}\\
    The AM classifier stores $n_{cl}$ binary prototype hypervectors of dimension $d$. 
    \item \textbf{HD encoder} \\ 
    The encoder requires $n_b$ frequency band hypervectors.
    However, it would be sufficient to store only one seed hypervector and compute (rematerialize) the remaining hypervectors at run time using e.g., a cellular automata~\cite{Yilmaz2015}. 
    \item \textbf{Thermometer/Gray code}\\
    Thermometer encoding does not need to store any dictionary since the code is deterministic. 
    It simply maps a level \textit{i} to \textit{i} number of 1's. 
    The same can be done for the Gray code as well when assuming the initial seed to be all zero and follow a specific rule which bits to change between levels. 
    \item \textbf{Random projection} \\
    The same projection matrix is shared among all frequency bands. 
    Therefore, only one projection matrix of dimension $n_R \times d$ is stored.
    Two bits are sufficient to represent an component in the random projections matrix as it consist only three different values. 
    Due to the sparsity of the random projection matrix, i.e. 90\% of the values are zero, the encoding could be optimized to 1.1 bits using simple Huffman coding. 
    \item \textbf{Learned projection} \\
    In contrast to the random projection the learned projection matrix is dense. 
    However, our evaluations show that representing the learned projection with \textit{float8} is sufficient to achieve the same classification accuracy as with \textit{float32}. 
    Similarly, the same learned projection matrix is shared among all frequency bands. 
\end{itemize}

\begin{figure*}[htbp]
  \subfloat[4-class MI]{\label{fig:sub_hd_map_results_a}{\includegraphics[width=0.475\textwidth]{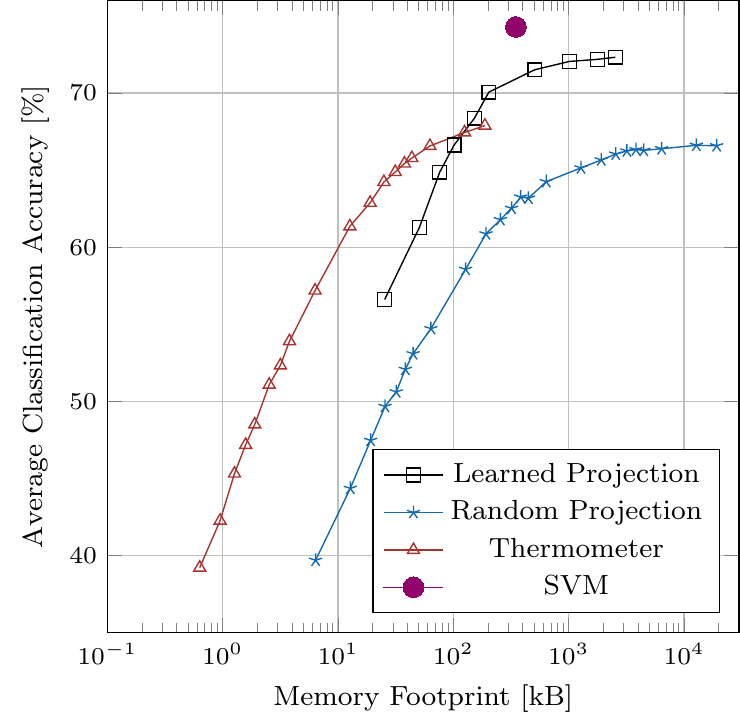}}}\hfill
\subfloat[3-class MI]{\label{fig:sub_hd_map_results_b}{\includegraphics[width=0.475\textwidth]{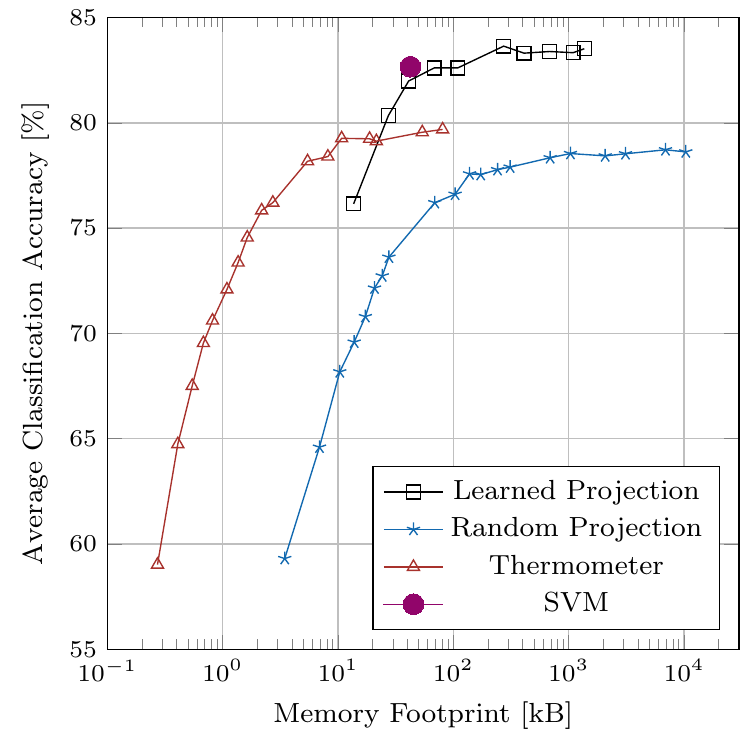}}}
\caption{Classification accuracy depending on the memory footprint of different classifier configurations. The memory footprint increases when using a higher dimension $d$ of the HD classifier.}
\label{fig:mem_footprint}
\end{figure*}

Fig. \ref{fig:mem_footprint} compares the average classification accuracy with the memory footprint on both datasets. 
Generally speaking, most of the configurations need less than 10\,MB memory to be stored on the device.
The random projections require the largest memory footprint on both datasets, even tough the encoding was set to two bits per matrix component.  
The flexibility of HD classifiers allows them to show graceful accuracy degradation with low memory footprint.
On the 3-class dataset, however, the memory footprint of the learned projection crosses the linear SVM at 40\,kB. 
On both datasets, the thermometer codes show the lowest memory footprint.

\begin{table*}
\vspace{0.05in}
\centering
\caption{Memory footprint calculation of the classifier modules in bits.}
\label{tab:memory}
\begin{tabular}{ll}
\cmidrule(r){1-2} 
SVM  & $64 \cdot n_{cl} \cdot n_R \cdot n_b$\\
\cmidrule{1-1} \cmidrule{2-2}
AM classifier &  $n_{cl} \cdot d$\\
HD encoder &  $d$\\
Thermometer/Gray & $0$  \\
Random projection &  $2 \cdot n_R \cdot d$\\
Learned projection & $8 \cdot n_R \cdot d$\\

\cmidrule(r){1-2}  
\end{tabular}
\end{table*}

\subsection{TX2 Board Measurements}
The HD classifier and the linear SVM are compared in terms of run time and energy consumption during training and testing. 
The features have been precalculated in advance to compare only the performance between the classifiers. 
For SVM, we use the Liblinear~\cite{Liblinear2008} library that provides a highly efficient implementation of the linear SVM for parallel execution on CPUs. 
It is dedicated for large-scale features, and therefore, a perfect match to the multi-spectral Riemannian features. 
The linear SVM is compared with a simple HD classifier using thermometer encoding.
%
%
Both the learning and inference of the HD classifier are implemented in \textit{C} on the GPU with \textit{cuda} support.

The measurement procedure is to first set the module in sleep mode for two seconds and then load the training or testing features only of the first subject of the dataset. 
Afterwards, the testing or training on the first subject is repeated up to 1,000$\times$ to get an averaged time and power estimation. 
The operation time is measured exclusively during inference or training. 
The consumed energy, however, is computed from two seconds to the end of the program by numerical integration of the power measurement. 
When using the GPU, the time and power consumption for transferring the model and features from the CPU to the GPU is included in every iteration. 
%

%
%


Table~\ref{tab:power} summarizes the power and time measurements on the NVIDIA TX2 board. 
All results are reported per trial. 
The energy consumption is estimated by numerically integrating the measurement of the MAIN power line. 
On both datasets and in both power modes, the HD classifier consumes less time and energy in training than the linear SVM. 
The maximum gain of 89.9$\times$ between HD and SVM training time is achieved on the 4-class dataset in MAX-Q mode, whereas the minimum gain of 9.67$\times$ is achieved on the 3-class dataset in MAX-N mode.
As a result, the HD classifier consumes less energy in training as well by a maximum of 58$\times$. 

In inference on the 4-class dataset, the HD classifier outperforms the linear SVM most of the time, but not significantly. 
On the 3-class dataset, however, the linear SVM outperforms the HD classifier during inference in all aspects and power modes. 
It is maximally 3.5$\times$ faster and consumes 3.9$\times$ less energy. 
The speed up of the SVM from the 4-class to the 3-class dataset is clearly evident due to the reduced feature dimension (10,879 vs. 1,768). 
The HD classifier however does not show the same benefits which may be due to the inefficient implementation of the AM.   
This issue can be addressed by an embedded processor supporting bitwise operations which demonstrated to increase the speed of inference by 3.4$\times$~\cite{DAc18}.

\begin{table*}
\vspace{0.05in}
\centering
\caption{Time and energy measurements per trial on NVIDIA TX2 board. The measurements are executed in both power optimized (MAX-Q) and performance optimized (MAX-N) modes. The best results are marked with bold.}
\label{tab:power}
\begin{tabular}{llrrrr}
\cmidrule(r){1-6} 
& & \multicolumn{2}{c}{4-class MI$^{\textrm{(a)}}$} & \multicolumn{2}{c}{3-class MI$^{\textrm{(b)}}$} \\
 \cmidrule(r){3-4}  \cmidrule(r){5-6} 
& & SVM & HD & SVM & HD\\
 \cmidrule(r){1-2} \cmidrule(r){3-3} \cmidrule(r){4-4} \cmidrule(r){5-5} \cmidrule(r){6-6} 
 & Classification Accuracy [\%] &\textbf{ 74.29} & 67.09 & \textbf{82.67} & 79.56\\
\cmidrule(r){1-2} \cmidrule(r){3-3} \cmidrule(r){4-4} \cmidrule(r){5-5} \cmidrule(r){6-6}
&Training Time [ms] &10.988 &\textbf{0.122}  &1.855 & \textbf{0.069}\\
MAX-Q & Inference Time [ms] & 0.182  &\textbf{0.108}  &\textbf{0.038} &0.094 \\
& Training Energy [mJ] &24.769 & \textbf{0.422} &4.721 & \textbf{0.212} \\
& Inference Energy [mJ] &0.437  &\textbf{0.362} &\textbf{0.098} &0.264 \\ 
\cmidrule(r){1-2} \cmidrule(r){3-3}\cmidrule(r){4-4}\cmidrule(r){5-5}\cmidrule(r){6-6}
 &Training Time [ms]  &4.603 &\textbf{0.119}  & 0.689 & \textbf{0.071} \\
MAX-N & Inference Time [ms] &\textbf{0.068}  & 0.078 & \textbf{0.016} & 0.058 \\
& Training Energy [mJ] &30.769  &\textbf{0.508} & 5.079 &\textbf{0.254} \\ 
& Inference Energy [mJ] &0.455  &\textbf{0.440} &\textbf{0.092} & 0.356\\


\cmidrule(r){1-6}  
\end{tabular}
\end{table*}

\section{Discussion}\label{sec:disc}
We have started by ad hod encoding of EEG signals using quantization based thermometer and Gray embedding as well as random projection.
They all showed moderate average classification accuracy on both datasets.
However encoder using thermometer and Gray coding demands an order of magnitude lower memory footprint than random projection at the same accuracy (see Fig.~\ref{fig:mem_footprint}). 
Further evaluation on the TX2 board showed that HD classifier with thermometer embedding is an excellent candidate when time and energy constraints are tight, and a moderate classification accuracy is sufficient.  

%
To the best of our knowledge, for the first time, we have shown that binary HD encoding can be coupled with binary neural networks for embedding real-valued features.
The encoder has been trained end-to-end to represent EEG signals with binary hypervectors.  
The training of projections forces the embedded hypervectors to be holographic, which further improved the performance both in terms of classification accuracy and required dimensionality of the embedding: learned projections with $d$=400 achieve the same classification accuracy as the linear SVM on the 3-class MI dataset. 
The encoder is capable to represent 1768 features of \textit{double64} with only 400 bits corresponding to a compression ratio of 283. 

Learned projections represent the features much more effectively than random projections because they are trained end-to-end to produce quasi-orthogonal hypervectors. 
This insight is supported by an application in image compression~\cite{KellyMatthewA2013}, where the quality increases with hypervector representation having uncorrelated components.

\section{Conclusion}\label{sec:conclusion}
In this paper, we propose a set of embedding methods to represent and classify EEG signals with binary hypervectors for an application of MI-based BCIs.
In addition to well established embeddings such as thermometer/Gray coding and random projections, we examine a novel encoding method which aims to learn projections with end-to-end training. 
Experimental results on two different MI datasets demonstrate that thermometer and Gray embeddings achieve up to 89.9$\times$ faster training and reach similar accuracy to random projections, however, when using learned projections the average accuracy increases to 84.22\% on the 3-class dataset, which is 1.55\% higher than state-of-the-art linear SVM, and 72.54\% on the 4-class dataset.

\section*{Acknowledgements}
Support was received from the ETHZ Postdoctoral Fellowship Program, the Marie Curie Actions for People COFUND Program, and the Hasler Foundation.




\bibliographystyle{model1-num-names}
\bibliography{bibliography.bib}







\end{document}